\newcommand{\circled}[1]{%
  \tikz[baseline=(char.base)]{
    \node[shape=circle,fill=black,draw,inner sep=1pt, font=\fontsize{9pt}{9pt}\selectfont] (char) {\textcolor{white}{#1}};%
  }%
}

\newcounter{romanCounter}
\documentclass[9pt, conference]{IEEEtran}
\IEEEoverridecommandlockouts

\usepackage{cite}
\usepackage{amsmath,amssymb,amsfonts}
\usepackage{algorithm}
\usepackage{algorithmicx}

\usepackage{algpseudocode}

\usepackage{graphicx}
\usepackage{textcomp}
\usepackage{xcolor}

\usepackage{booktabs}

\usepackage{graphicx}

\usepackage{colortbl}
\usepackage{tabularx}
\usepackage{multirow}

\usepackage{CJKutf8}
\usepackage{pifont}
\usepackage{float} 
\usepackage{colortbl} 
\usepackage{diagbox} 

\usepackage{threeparttable}

\usepackage{tikz}
\usepackage{xcolor} 

\usepackage{booktabs}
\usepackage{multirow}
\usepackage{makecell}
\usepackage{colortbl}
\usepackage{xcolor}
\usepackage{graphicx}
\usepackage{subcaption}

\usepackage{url}
\usepackage[hidelinks]{hyperref}

\definecolor{deepgreen}{rgb}{0.0, 0.5, 0.0}

\def\BibTeX{{\rm B\kern-.05em{\sc i\kern-.025em b}\kern-.08em
    T\kern-.1667em\lower.7ex\hbox{E}\kern-.125emX}}
    
\begin{document}

\title{Insights from Rights and Wrongs: A Large Language Model for Solving Assertion Failures in RTL Design
}

\author{
    \IEEEauthorblockN{
        Jie Zhou$^{1,2}$, Youshu Ji$^2$, Ning Wang$^3$, Yuchen Hu$^{1,2}$, Xinyao Jiao$^{1,2}$, Bingkun Yao$^3$, \\
        Xinwei Fang$^4$, 
        Shuai Zhao$^5$, Nan Guan$^3$, Zhe Jiang$^{1,2}$
    }
    \IEEEauthorblockA{
        $^1$School of Integrated Circuits, Southeast University, China \\
        $^2$National Center of Technology Innovation for EDA, China\\
        $^3$Department of Computer Science, City University of Hong Kong, Hong Kong\\
        $^4$Department of Computer Science, University of York, UK \\
        $^5$Department of Computer Science, Sun Yat-sen University, China
    }
}

\maketitle

\thispagestyle{plain}
\pagestyle{plain}

\begin{abstract}

SystemVerilog Assertions (SVAs) are essential for verifying Register Transfer Level (RTL) designs, as they can be embedded into key functional paths to detect unintended behaviours. 
During simulation, assertion failures occur when the design's behaviour deviates from expectations. 
Solving these failures, i.e., identifying and fixing the issues causing the deviation, requires analysing complex logical and timing relationships between multiple signals. This process heavily relies on human expertise, and there is currently no automatic tool available to assist with it.
Here, we present AssertSolver, an open-source Large Language Model (LLM) specifically designed for solving assertion failures. By leveraging synthetic training data and learning from error responses to challenging cases, 
AssertSolver achieves a bug-fixing pass@1 metric of 88.54\% on our testbench, significantly outperforming OpenAI's o1-preview by up to 11.97\%. 
We release our model and testbench for public access to encourage further research: \textbf{\url{https://github.com/SEU-ACAL/reproduce-AssertSolver-DAC-25}}.

\end{abstract}

\section{Introduction}

Functional verification is a crucial step in the modern Electronic Design Automation (EDA) process, ensuring that designs meet their specifications and perform correctly as intended \cite{wang2009electronic}, thereby mitigating the costly risks associated with silicon failures \cite{rajendran2015detecting, witharana2022survey}. SystemVerilog Assertions (SVAs), as one of the key methods in functional verification \cite{ranjan2009beyond}, capture potential errors in Register Transfer Level (RTL) designs by defining logical conditions and timing requirements. Unlike stimulus-based verification methods (e.g., using testbenches), SVAs not only facilitate stimulus-triggered checks but also enable formal verification \cite{seligman2023formal}. Formal verification, which mathematically proves the consistency of the design under test (DUT) with the behaviour specified by the SVAs, provides higher functional coverage and addresses boundary conditions that stimulus-based methods may often overlook.

\begin{figure}[t]
    \centering
    \scalebox{1.0}{\includegraphics[width=\columnwidth]{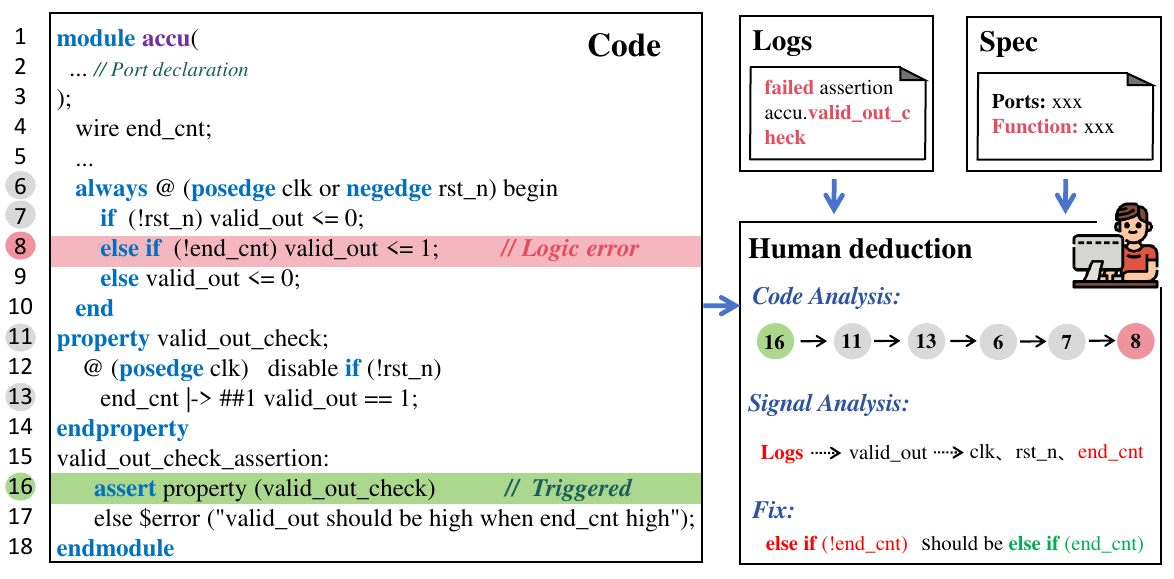}}
    \vspace{-15pt}
    \caption{The process of solving assertion failures in RTL verification. Engineers analyse the design specification and code signals to identify the root causes of assertion failures and implement fixes.}  
    \vspace{-15pt}
    \label{fig:intro}
\end{figure}

Despite the significant advantages of SVAs and progress in automating their generation~\cite{maddala2024laag, miftah2024assert, fang2024assertllm, pulavarthi2024assertionbench, orenes2023using, sun2023towards}, automatically solving assertion failures remains challenging. Assertion failures occur when the design exhibits unexpected behaviour during simulation.
Effectively identifying and fixing these failures still relies heavily on human expertise to analyse intricate logical dependencies and timing relationships between multiple signals, which is time-consuming and labour-intensive. Fig. \ref{fig:intro} illustrates 
that verification engineers must have an in-depth understanding of the design’s functional intent and carefully deduce the causes of complex assertion failures.

\begin{figure*}[t]
\includegraphics[width=1\textwidth]{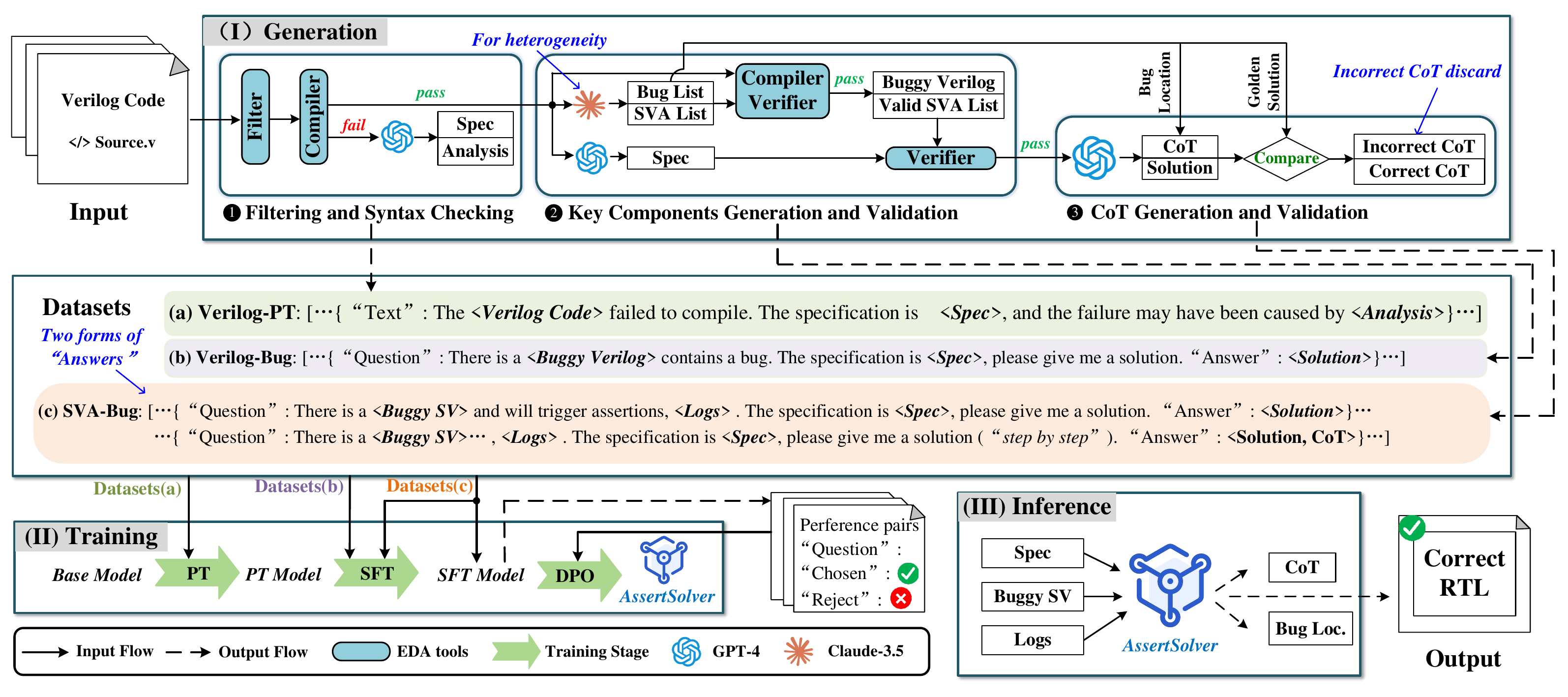}
\vspace{-15pt}
\caption{AssertSolver overview: \textbf{(I)} shows the process for augmenting the training data. \textbf{(II)} describes the training strategy, consisting of pretraining (PT), Supervised Fine-Tuning (SFT), and Direct Preference Optimisation (DPO) which is used to guide learning from error responses to challenging cases. In the inference phase \textbf{(III)}, AssertSolver locates and fixes the bug based on the specification (Spec), buggy SystemVerilog (SV) code, and logs, while also providing bug location (Bug Loc.) and an explanation (CoT).}
\vspace{-10pt}
\label{fig:data}
\end{figure*}

Large Language Models (LLMs) have demonstrated significant capabilities in navigating through complex ideas, automating key steps in hardware design. They have improved a range of tasks, from code generation~\cite{liu2023verilogeval, thakur2024verigen, wang2024large, ho2024verilogcoder, thakur2023autochip} to design verification~\cite{xu2024meic, fu2023llm4sechw, yao2024hdldebugger}, offering potential for solving assertion failures.
However, while state-of-the-art (SOTA) LLMs such as OpenAI's GPT-4 and o1-preview demonstrate competence in such task, their performance is often subject to unnoticed changes, and they lack the ability to be retrained or fine-tuned by users to incorporate new data or information.

To address this limitation, we introduce an open-source LLM, AssertSolver, specifically designed for automatically solving assertion failures. As shown in Fig. \ref{fig:data}, AssertSolver benefits from an innovative data augmentation method that enhances the representation of various assertion failure scenarios in the training dataset, which were previously underrepresented. AssertSolver is retrained and fine-tuned from the Deepseek-Coder-6.7b \cite{guo2024deepseek}, using the dataset that incorporates not only significantly enhanced examples but also error responses to challenging cases. By learning from these ``right'' and ``wrong'' examples, 
AssertSolver demonstrates up to a 11.97\% improvement in solving assertion failures compared to the OpenAI's o1-preview. 

The main contributions of this paper are:
\begin{itemize}

\item Development of AssertSolver, an open-source domain-specific LLM fine-tuned for solving assertion failures, which we have made publicly available for early adoption;

\item Implementation of a data augmentation method that automatically enrich scarce training dataset for assertion failures; 

\item Adoption of a new training strategy that improves learning from error responses to challenging cases;

\item Publication of openly accessible benchmark for solving assertion failures, featuring over 900 instances across various bug types.

\end{itemize}

The rest of the paper is organised as follows: Section \ref{sec:Automatic dataset generation} introduces the data augmentation approach and the preparation of the training dataset. Section \ref{sec:training} provides a detailed description of our training strategies. Section \ref{sec:Setup} outlines the research questions and the experimental setup. Section \ref{sec:Results} presents a comprehensive evaluation to answer the research questions. Section \ref{sec:Conclusion} concludes the paper.

\section{Data Augmentation}
\label{sec:Automatic dataset generation}

To effectively train an LLM for solving assertion failures, it is essential to have access to a comprehensive training dataset. While current datasets contain a significant number of Verilog code samples suitable for preliminary pretraining, they lack key components required for fine-tuning the model to this specific task. These components include design specifications, SVAs, details of bugs that could invalidate these assertions, and their corresponding verification outcomes. To address this gap, we propose a method that integrates LLMs and EDA tools to augment existing open-source Verilog code, thereby generating the necessary components for solving assertion failures. Additionally, to help users understand the model's problem-solving process, we have incorporated a Chain of Thought (CoT) \cite{wei2022chain} in the dataset. This inclusion is designed to enhance the transparency of the model's decision-making process. In this work, we augmented an open-source dataset~\cite{thakur2023benchmarking} with 108,971 Verilog code samples from Hugging Face in the following 3 stages, as shown in Fig. \ref{fig:data}-\textbf{(I)}.

\textbf{Stage \circled{1}: Filtering and Syntax Checking.} Given the Verilog code samples from~\cite{thakur2023benchmarking}, we employed scripts to filter out code that exhibited certain undesirable characteristics. Specifically, we excluded code samples based on the following criteria: 
\begin{enumerate}
    \item Incomplete code that lacks either the `\textit{module}' or `\textit{endmodule}';
    \item Code comprising only initialisation or assignment statements, with no functional logic;
    \item Duplicated code segments.
\end{enumerate}

Then, each remaining piece of code underwent a syntax check through the Icarus Verilog compiler \cite{williams2002icarus}. Code with detectable syntax errors (e.g., those that failed the compilation), were excluded from the following process, preventing their advancement to the following stages where assertions could be triggered. However, these incorrect code samples were not discarded; instead, they are included in the \textbf{Verilog-PT} dataset as they offer valuable structural insights into Verilog code. After the syntax check, GPT-4 was employed to generate a design specification (Spec) for all code samples, but only those that failed compilation received further analysis. The analysis aims to provide an explanation for the causes of the syntax errors, contributing to the formation of the \textbf{Verilog-PT} dataset, as shown in Fig. \ref{fig:data} dataset (a). The \textbf{Verlog-PT} dataset, containing 22,646 entries, which is utilised in section \ref{sub:PT}. 

\textbf{Stage \circled{2}: Key Components Generation and Validation.} 
For successfully compiled code, we employed Claude-3.5 to generate random bugs and SVAs. 
The use of Claude-3.5, as opposed to GPT-4, was intended to leverage the heterogeneous design of these models, thereby helping GPT-4 avoid falling into error traps during self-validation in \textbf{Stage \circled{3}}. 

To mitigate the impact caused by the hallucinations of LLMs\cite{perkovic2024hallucinations, reddy2024hallucinations,tonmoy2024comprehensive}, we utilised EDA tools at this stage to validate the correctness of the generated components. We employed a two-step verification process by integrating the compiler with the verifier, SymbiYosys \cite{wolf2022symbiyosys}. Each generated SVA was inserted into the corresponding Verilog code and verified using SymbiYosys to ensure the SVAs' validity. Furthermore, we employed the compiler again to identify and eliminate syntax errors introduced during the random bug generation process. The remaining bugs were injected individually into the Verilog code, and each modified version was then verified with the corresponding SVAs using SymbiYosys to ensure that the bug-SVA pair caused assertion failures and to generate logs.

Approximately 90\% of the bugs and SVA pairs associated with assertion failures were selected and processed in \textbf{Stage \circled{3}}. The remaining 10\% is reserved for testing, as detailed in Section \ref{sbsc:Benchmarks}. This selection process is structured to ensure that the training and test datasets are completely separate, involving the following steps:  
\begin{enumerate}  
    \item Organise the buggy Verilog code into categories based on the length of the code, with bins defined as: \((0, 50]\), \((50, 100]\), \((100, 150]\), \((150, 200]\), and \((200, +\infty)\);  
    \item Enumerate the unique module names within each bin;
    \item Uniformly select 90\% of the module names and their corresponding buggy SystemVerilog (SV) code, Spec, and logs for inclusion in the training dataset. 
\end{enumerate}   

Bugs that did not cause assertion failures, potentially due to insufficient SVA coverage, were retained as they represent functional issues in the original Verilog code. These bugs, together with the Spec, Verilog code, and solution, were organised into the \textbf{Verilog-Bug} dataset. This dataset, as shown in Fig. \ref{fig:data} dataset (b), contains 36,650 entries and is used in section \ref{sub:SFT} to enhance the model's understanding of the verification process.  

\textbf{Stage \circled{3}: CoT Generation and Validation}. 
To enhance the transparency of our model, we incorporated CoTs into the training data. This integration involves using GPT-4, where we provided Spec, buggy SV code, logs, and the bug location. The task for GPT-4 is to generate a CoT that articulates its reasoning behind identifying the erroneous code and suggesting correction. 

Subsequently, we used a script to validate these CoTs by comparing GPT-4's output with the `golden solution' obtained from Stage \circled{2}. The `golden solution' was derived from the initial buggy code and its correct counterpart. If GPT-4 identified an error and proposed correction align with it, we consider the CoT to be correct.

In total, approximately 74.55\% of the generated CoTs were identified to be correct. Depending on the correctness of these CoTs, we organised two types of entries into the \textbf{SVA-Bug} dataset, which totals 7,842 entries as shown in Fig. \ref{fig:data} dataset (c). For entries where the CoT is correct, the `Question' section includes the phrase `\textit{step by step}', and the `Answer' details both the buggy line and its corrected code, along with the CoT. Otherwise, the `Answer' only includes the buggy line and the correct code. This dataset, designed specifically to equip the model with the ability to solve assertion failures and explain the reasoning process, is utilised in section \ref{sub:SFT}.

\section{training strategy}
\label{sec:training}

\begin{table*}[t]
    \centering
    \caption{Bug types leading to assertion failures and examples} 
    \begin{tabular}{p{1.5cm} >{\raggedright\arraybackslash}p{6.5cm} >{\raggedright\arraybackslash}p{3cm} >{\raggedright\arraybackslash}p{3cm} >{\raggedright\arraybackslash}p{2cm}}
    \toprule
        \rowcolor[gray]{0.9}
        \textbf{Type} & \textbf{Description} & \textbf{Expected Form} & \textbf{Unexpected Form} & \textbf{Assertion}$^\dagger$ \\ \midrule
        \textbf{Direct} & Bug signal appears directly in the assertion. & 
        \cellcolor{green!15}out \texttt{<=} \textcolor{deepgreen}{\textbf{in;}} & 
        \cellcolor{red!15}out \texttt{<=} \textcolor{red}{\textbf{in + 1;}} &
        assert(out == \textcolor{deepgreen}{\textbf{in}})
        \\ \midrule

        \textbf{Indirect} & Bug signal does not appear directly in the assertion. & 
        \cellcolor{green!15}\makecell[l]{\textcolor{deepgreen}{\textbf{temp \texttt{<=} in;}}\\ out \texttt{<=} temp;} & 
        \cellcolor{red!15}\makecell[l]{\textcolor{red}{\textbf{temp \texttt{<=} in + 1;}}\\ out \texttt{<=} temp;} &
        assert(out == in) \\ \midrule

        \textbf{Var} & Incorrect variable name or type. & \cellcolor{green!15}out = \textcolor{deepgreen}{\textbf{in}}; & \cellcolor{red!15}out = \textcolor{red}{\textbf{input}}; & -- \\ \midrule
        
        \textbf{Value} & Incorrect variable values, constants, or signal bit widths. & \cellcolor{green!15}out = \textcolor{deepgreen}{\textbf{4'b1010;}} & \cellcolor{red!15}out = \textcolor{red}{\textbf{4'b1110;}} & -- \\ \midrule
        
        \textbf{Op} & Misuse of operators. & \cellcolor{green!15}out = a \textcolor{deepgreen}{\textbf{\texttt{|}}} b; & \cellcolor{red!15}out = a \textcolor{red}{\textbf{\&}} b; & -- \\ \midrule

        \textbf{Cond} & Bug in conditional statement (e.g., \textit{if}, \textit{always}).  & 
        \cellcolor{green!15}\textcolor{deepgreen}{\textbf{if (valid)}} out \texttt{<=} in; & 
        \cellcolor{red!15}\textcolor{red}{\textbf{if (!valid)}} out \texttt{<=} in; & -- \\ \midrule
        \textbf{Non\_cond} & Bug unrelated to conditional statements. & \cellcolor{green!15}if (valid) \textcolor{deepgreen}{\textbf{out \texttt{<=} in}}; & 
        \cellcolor{red!15}if (valid) \textcolor{red}{\textbf{out \texttt{<=} input;}} & -- \\ \bottomrule
        \\[-2.5mm]
        \multicolumn{5}{l}{$^\dagger$ \scriptsize The distinction between \textit{Direct} and \textit{Indirect} type depends on whether the assertion failure is caused by the directly protected signal. Other types of bugs may cause assertion}\\
        \multicolumn{5}{l}{\scriptsize failures but are not necessarily directly reflected in the assertions.}
    \end{tabular}
    \label{tab:Common assertion failures categories and examples}
\end{table*}

\begin{table}[t]
\centering
\caption{Distribution of SVA-Bug and SVA-Eval across code length intervals and bug types (counts of instances)}
\label{table:CombinedDistribution}
\renewcommand{\arraystretch}{1.2} 
\begin{threeparttable} 
\resizebox{\columnwidth}{!}{
\begin{tabular}{c c c c c c} 
\toprule
\rowcolor[gray]{0.9}
\textbf{Length Interval} & (0, 50] & (50, 100] & (100, 150] & (150, 200] & (200, +$\infty$) \\ 
\midrule
SVA-Bug & 3400 & 2444 & 921 & 431 & 646 \\ 
SVA-Eval & 431 & 260 & 102 & 58 & 64 \\ 
\midrule[1pt]
\rowcolor[gray]{0.9}
\textbf{Bug Type} & Direct & Indirect & Var & Value & Op \\ 
\midrule
SVA-Bug & 5478 & 2364 & 546 & 5104 & 2254 \\ 
SVA-Eval & 615 & 300 & 47 & 601 & 274 \\ 
\midrule
\rowcolor[gray]{0.9}
\textbf{Bug Type} & Cond & Non\_cond & -- & -- & -- \\ 
\midrule
SVA-Bug & 1573 & 6269 & -- & -- & -- \\ 
SVA-Eval & 204 & 711 & -- & -- & -- \\ 
\bottomrule
\end{tabular}
}
\vspace{-15pt}
\end{threeparttable} 
\end{table}

To maximise the potential of the dataset generated in Section \ref{sec:Automatic dataset generation}, we implemented a tailored training strategy for AssertSolver, as shown in Fig. \ref{fig:data}-\textbf{(II)}. Our strategy started with a foundational \textbf{pretraining} \textbf{(PT)} phase using the base model, Deepseek-Coder-6.7b, on the \textbf{Verilog-PT} dataset. This step was to strengthen the model's understanding of Verilog code constructs and design specifications. Following this, we applied \textbf{Supervised Fine-Tuning (SFT)} to fine-tune the AssertSolver on the \textbf{SVA-Bug} and \textbf{Verilog-Bug} datasets. This step was designed to equip the model with the necessary skills for solving assertion failures, and to expand its capability for generalising across related Verilog debugging tasks. To further refine the model performance, we revisited unresolved assertion failure cases from the SFT phase, particularly those with incorrect responses, and employed \textbf{Direct Preference Optimisation (DPO)} to enable our model to learn from these error responses to challenging cases.

\subsection{Pretraining}

\label{sub:PT}

\textcolor{red}{} 
Pretraining is essential for LLMs, especially when preparing them to handle specialised languages such as Verilog and SystemVerilog. It is important to infuse domain-specific knowledge during this stage, which forms a robust base of understanding before any targeted fine-tuning starts. 
Recent research~\cite{yao2024location, wang2024large} supports that continual pretraining on domain-specific dataset, including unlabeled Verilog code and syntactically similar language like C/C++ code, can substantially boosts the base model’s understanding of hardware description languages (HDLs) and improve its performance in downstream tasks. 
In line with these insights, we implemented continual pretraining with the\textbf{ Verilog-PT} dataset, comprising the Verilog code that failed in compilation along with its corresponding specifications and analyses of compilation failures. The base model, Deepseek-Coder-6.7b, has been preliminarily trained on a large programming corpus and is lightweight, making it ideal for this application. This focused pretraining strategy is essential as the timing and concurrency property in HDLs differ significantly from software programming paradigms.

During pretraining, each sample \( x^{(i)} \) in the \textbf{Verilog-PT} dataset \( D_{\text{PT}} = \{ x^{(i)} \}_{i=1}^{N} \) is treated as a sequence of tokens. 
These tokens serve as the basic units in natural language processing tasks, allowing LLMs to leverage reasoning over them for next-token predication, which in turn facilitates text generation
\cite{mielke2021between,qi2024next,sennrich2015neural, shibata1999byte}. The sequence for each sample is expressed as \( x^{(i)} = (w_1^{(i)}, w_2^{(i)}, \ldots, w_{T^{(i)}}^{(i)}) \), where \( w_t^{(i)} \) denotes the \( t \)-th token and \( T^{(i)} \) the total number of tokens in the sequence. The training objective in this stage focuses on minimising the negative log-likelihood loss across the dataset: 

\begin{equation*}
\small
L_{\text{PT}}\left(\theta\right) = -\sum_{i=1}^{N} \sum_{t=1}^{T^{(i)}} \log P(w_t^{(i)} \mid \text{context}_t^{(i)}; \theta)
\end{equation*}

Here \( \text{context}_t^{(i)} \) refers to the series of preceding tokens which serve as the basis for predicting \( w_t^{(i)} \), and \( P(w_t^{(i)} \mid \text{context}_t^{(i)}; \theta) \) represents the probability of predicting \( t \)-th token based on this context, as determined by the model's parameter \( \theta \). This pretraining lays the groundwork for the subsequent fine-tuning process and equips the model capable of handling tasks within the hardware design domain. 

\subsection{ Supervised Fine-Tuning (SFT)}
\label{sub:SFT}

Following the unsupervised pretraining phase, where the model primary learning was to predict the next token, Supervised Fine-Tuning (SFT) aims to shift the model’s capabilities. This shift moves the focus from mere text continuation to solving specific question-answering challenges presented by assertion failures, which require precise and supervised responses.

To this end, we used the \textbf{SVA-Bug} dataset, which includes Spec, buggy SV code and logs. These elements are organised into the model's input \( x \) , as shown in the `Question' section in Fig. \ref{fig:data} dataset (c). The model output the `Answer' \( y \) must include, at a minimum, the bug line snippet and the corresponding correct code.  Additionally, if the CoT is verified as correct in \textbf{Stage \circled{3}} of section \ref{sec:Automatic dataset generation}, it is also included in \( y \), enhancing the answer with detailed reasoning steps, marked by the keyword `step by step' in \( x \). Furthermore, we integrated the \textbf{Verilog-Bug} dataset as an auxiliary task to further enrich the training data with a broader spectrum of Verilog debugging scenarios. The data pairs \( \langle x, y \rangle \) from this dataset are structured to provide the model with both the buggy Verilog code in the input `Question' and the repair plan in the `Answer', which lists the buggy line and alongside the corrected version, as shown in Fig. \ref{fig:data} dataset (b). This combination of datasets in the SFT process ensures a comprehensive learning experience for the model.

The objective of SFT is to refine the model’s ability to predict the next token in \( y^{(i)} \) based on the contextual interplay between the input \( x^{(i)} \) and the sequence of previously generated tokens \( y_{<t}^{(i)} \). This approach is designed to train the model in recognising and replicating the correct relationships between given `Question' and `Answer':

\begin{equation*}
\small
L_{\text{SFT}}\left(\theta\right) = -\sum_{i=1}^{N} \sum_{t=1}^{T_y^{(i)}} \log P(y_t^{(i)} \mid y_{<t}^{(i)}, x^{(i)}; \theta)
\end{equation*}

where \( y_t^{(i)} \) denotes the \( t \)-th token in the output sequence \( y^{(i)} \), and \( y_{<t}^{(i)} \) represents the sequence of tokens preceding \( y_t^{(i)} \). The likelihood \( P(y_t^{(i)} \mid y_{<t}^{(i)}, x^{(i)}; \theta) \) indicates the probability of predicting the token \( y_t^{(i)} \) given the \( y_{<t}^{(i)} \) and the input context \( x^{(i)} \), as governed by the model parameters \( \theta \). The SFT allows the model to produce the answer in the expected format and to develop a deeper understanding of the underlying hardware description language. 

\subsection{Learning from Error Responses to Challenging Cases}
\label{sub:Learn from Error Responses to Challenging Cases}

At the SFT stage, AssertSolver is primarily exposed to correct responses, which limits its ability to process or learn from errors - a critical aspect of human learning. As noted in research by \cite{mercer2008talk,reich2023overcome}, effective learning involves not only assimilating correct responses but also reflecting on and learning from mistakes to prevent future errors. Current training paradigms often overlook or discard erroneous data. Inspired by recent research \cite{tong2024can, chen2023gaining, an2023learning}, we propose to equip AssertSolver with the ability to learn from its errors, thus enhancing its decision-making process.

To facilitate this, we evaluate the SFT model using all samples within the \textbf{SVA-Bug} dataset. Each sample, as illustrated in Fig. \ref{fig:data} dataset (c), includes a `Question' section, which serves as the model’s input. For each input, the model generates 20 responses. Correctness is then evaluated by comparing the buggy line suggested by the model with the correct  `Answer' in the dataset. Samples yielding at least one incorrect response among these 20 outputs are categorised as \textit{challenging cases}, representing instances where the model struggles despite prior exposure to correct solutions. In each challenging case, the `Question' is denoted as  \( x \)  and the correct `Answer' as \( p \). The incorrect responses to \( x \) are denoted as \( n[k] \), where \( k < 20 \).

\begin{sloppypar}

We abstract them into triples: \( D_{\text{DPO}} = \{(x^{(i)}, p^{(i)}, n[k]^{(i)})\}_{i=1}^{N} \), where \( N \) denotes the number of \textit{challenging cases}.
For this preference dataset \( D_{\text{DPO}} \), the objective is to train the model to prioritise generating the correct response \( p^{(i)} \) over the error responses \( n[k]^{(i)} \) for each \( x^{(i)} \). This goal is achieved through the DPO loss function that encourages the model to maximise the probability of generating the correct response \( p^{(i)} \), while minimising the probability of generating the error response \( n[k]^{(i)} \):

\end{sloppypar}

\[
\small
\resizebox{\columnwidth}{!}{$L_{\text{DPO}} = -\mathbb{E}_{D} \left[ \log \sigma \left( \beta \log \frac{\pi_{\theta}(p^{(i)} | x^{(i)})}{\pi_{\text{ref}}(p^{(i)} | x^{(i)})} - \beta \log \frac{\pi_{\theta}(n[k]^{(i)} | x^{(i)})}{\pi_{\text{ref}}(n[k]^{(i)} | x^{(i)})} \right) \right]$}
\]

In this function, \( \pi_{\theta} \) and \( \pi_{\text{ref}} \) represent the current model and the reference model (SFT model, in this context). The terms \( \pi_{\theta}(p^{(i)} | x^{(i)}) \) and \( \pi_{\text{ref}}(p^{(i)} | x^{(i)}) \) denote the likelihood of generating the correct response \( p^{(i)} \) given input \( x^{(i)} \) under the respective models. Similarly, \( \pi_{\theta}(n[k]^{(i)} | x^{(i)}) \) and \( \pi_{\text{ref}}(n[k]^{(i)} | x^{(i)}) \) correspond to the probabilities of generating the \( k \)-th error response \( n[k]^{(i)} \) for the same input. The log-ratios of these probabilities quantify the divergence between the two models. The difference between the log-ratios captures the preference of \( \pi_{\theta} \) for generating the correct response \( p^{(i)} \) over the error response \( n[k]^{(i)} \), guiding the model producing correct answers. The scaling factor \( \beta \), set to 0.1, controls the weight of the log-ratio terms, and the sigmoid function \( \sigma \) maps the log-ratio values to a probability in the range \([0, 1]\), facilitating smooth learning while ensuring training stability. This approach allows AssertSolver not only to learn from errors but also to improve its response accuracy in challenging cases, leading to more robust decision-making capabilities.

\section{Evaluation}
\label{sec:Setup}
To evaluate the effectiveness of our trained model, we conduct extensive experiments designed to answer four key research questions. This section outlines the dataset property, benchmark and SOTA counterparts, evaluation metrics, and implementation details.

\subsection{Research Questions}
The evaluation is structured to investigate the following four research questions:
\begin{itemize}
    \item \textbf{RQ1:} How does the incorporation of learning from error responses influence the performance of the model as measured by metrics of \textit{pass@1} and \textit{pass@5}?
    \item \textbf{RQ2:} How does the effectiveness of our model in solving assertion failures compare to that of other SOTA LLMs or models of similar complexity?
    \item \textbf{RQ3:} How does the model’s performance vary when addressing randomly generated bugs versus human-crafted cases?
    \item \textbf{RQ4:} How is the model’s performance impacted by design variability, specifically in relation to different bug types and variations in code length?
\end{itemize}

\subsection{Dataset Property}
\label{sbsc:Classification}

The challenge of solving assertion failures varies considerably across different bug types and code lengths. 
Some categories are inherently more complex than others.
For example, timing-related bugs that do not directly trigger assertion failures require deep reasoning and analysis, making them more complex for verification engineers. Similarly, longer code lengths may increase the complexity of debugging, as they often involve more intricate logic and a higher potential for subtle errors. Recognising this, we classified the types of bugs leading to assertion failures, as shown in Table \ref{tab:Common assertion failures categories and examples}. We also analysed the code lengths and the number of instances falling into each identified bug category within our training and testing datasets, as illustrated in Table \ref{table:CombinedDistribution}. This classification is essential for evaluating the performance of LLMs across different categories, as it helps determine whether LLMs can effectively address bug types consistent with our expectations.

\subsection{New Open-Source Benchmark and SOTA Counterparts}
\label{sbsc:Benchmarks}

Given the lack of open-source benchmarks for evaluating tasks related to solving assertion failures, we have developed  a benchmark, \textbf{SVA-Eval}, to address this gap. This benchmark consists of 877 samples generated by LLMs (\textbf{SVA-Eval-Machine}), as described in Section \ref{sec:Automatic dataset generation}, and 38 manually curated samples (\textbf{SVA-Eval-Human}) derived from the RTLLM dataset \cite{lu2024rtllm}, ensuring both the scale of the dataset and the inclusion of real-world scenarios.
Each sample in \textbf{SVA-Eval} includes the Spec, buggy SV code, logs, and correct solutions, providing a comprehensive resource for evaluation. The dataset is publicly available on \textbf{\url{https://github.com/SEU-ACAL/reproduce-AssertSolver-DAC-25}} to support further research in this domain. 

For the comparison between AssertSolver and the current SOTA LLMs, we have chosen several commercially available closed-source models as benchmarks. These include Claude-3.5, GPT-4 \cite{achiam2023gpt}, and OpenAI’s latest o1-preview, all of which have demonstrated exceptional capabilities in RTL generation and debugging tasks. Furthermore, we have extended our comparative analysis to include open-source models such as CodeLlama-6.7b \cite{roziere2023code}, Llama-3.1-8b, and our base model, Deepseek-Coder-6.7b \cite{guo2024deepseek}. This inclusive approach aims to provide a comprehensive overview of AssertSolver’s performance across a spectrum of platforms and development environments.

\subsection{Evaluation Metrics}
\label{sbsc:Metrics}

\begin{table}[t]
\centering

\caption{Model performance as pass@k  (grey shading: the best performance across models).}
\label{tab:Pass@k under different training strategies.}
\setlength{\tabcolsep}{12pt}
\begin{tabular}{@{}lccc@{}}
\toprule
\textbf{Metric} & \textbf{Base Model} & \textbf{SFT Model} & \textbf{AssertSolver} \\
\midrule
\textit{Pass@1} & 4.35\% & 84.66\% & \colorbox{gray!25}{88.54\textbf{\%}} \\
\textit{Pass@5} & 15.62\% & \colorbox{gray!25}{91.64\textbf{\%}} & 90.00\% \\
\bottomrule
\end{tabular}
\end{table}

To evaluate the performance of our LLM in solving assertion failures, we employ the \textbf{\textit{pass@k}} metric, widely used in the evaluation performance of hardware designs \cite{thakur2023benchmarking,huang2024towards,liu2024rtlcoder}. This metric quantifies the effectiveness of LLMs by measuring their ability to generate correct solutions for each buggy SV code that causes assertion failures. For each instance, the model is provided with the buggy SV code alongside its corresponding specifications and logs, from which it generates \( n \) possible solutions. We then assess these solutions, deeming \( c \) of them effective if they successfully solve the assertion failure. This approach offers an unbiased estimate of the likelihood that at least one of the top \( k \) selections will address the problem, as shown by the following equation:

\begin{equation*}
\small
pass@k = \mathbb{E}_{\text{problems}} \left[ 1 - \frac{\binom{n - c}{k}}{\binom{n}{k}} \right]
\end{equation*}

In this study, we set \( n = 20 \) and \( k = \{1, 5\} \).  

\begin{itemize}
    \item \textbf{\textit{Pass@1}} assesses the model's accuracy and consistency by requiring the correct solution to be produced on the first attempt. An improvement in \textit{pass@1} suggests that the model is becoming more adept in identifying and responding accurately to the bugs, thus likely improving its precision for such tasks.
    \item \textbf{\textit{Pass@5}} evaluates whether model can provide at least one correct solution within five attempts. An increase in \textit{pass@5} indicates  that there is an enhancement in the model’s ability to generate diverse solutions, reflecting its flexibility in problem-solving.

\end{itemize}

\subsection{Implementation Details}
\textbf{\textit{Training.}} We fine-tuned the Deepseek-Coder-6.7b model using 8 A800-80G GPUs and accelerated the process with DeepSpeed ZeRO-3 \cite{rasley2020deepspeed}. We opted for full-parameter fine-tuning to achieve optimal performance and set an initial learning rate of \(10^{-4}\) for pretraining and SFT, incorporating a warm-up phase during the first 10\% of training steps.
% , along with a cosine learning rate schedule for smoother convergence. 
For DPO, a lower learning rate of \(10^{-6}\) was used, as it focuses on learning the difference between correct and incorrect answers, rather than directly optimizing for explicit answers.

\textit{\textbf{Inference.}} 
We combined the Spec, buggy SV code, and logs from the benchmark, requiring LLMs to return responses in a JSON format with a candidate buggy line, suggested fix, and CoT (Fig. \ref{fig:data}-\textbf{(III)}). In experiments, we found open-source models often deviated from the prompt format, so we iteratively refined prompts until \( n = 20 \) JSON responses were generated per assertion failure case for \textit{pass@k} evaluation. The temperature was 0.2 for consistent yet diverse outputs, except for o1-preview, which has a fixed temperature that cannot be adjusted through the application programming interface. 
\section{Results and Analysis } 
\label{sec:Results}

\begin{figure}[t]
    \centering
    \scalebox{1.0}{\includegraphics[width=\columnwidth]{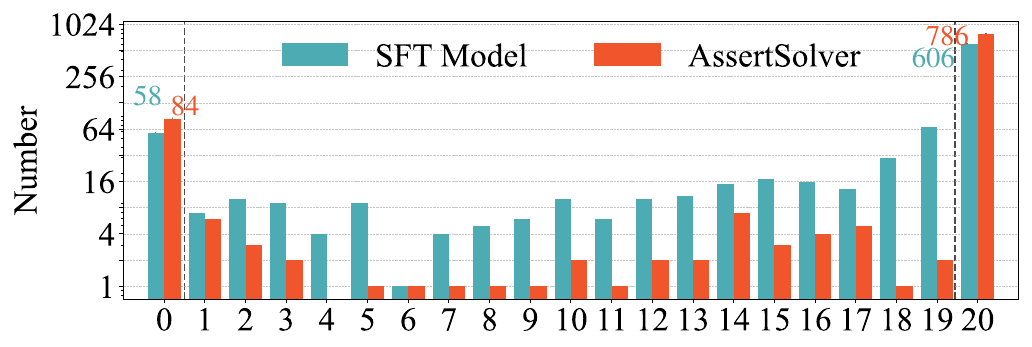}}
    \vspace{-10pt}
    \caption{Histogram of correct answers across 20 responses. (x-axis: \textit{c} (number of correct solutions in 20 responses))}
    \vspace{-15pt}
    \label{fig:Frequency distribution of correct answers across 20 responses.}

\end{figure}

\label{sbsc:Insights from rights and wrongs}
\noindent{\textbf{RQ1}}: To answer this question, we compared three models: the base model, the SFT model, and AssertSolver, using the \textbf{SVA-Eval} dataset. As shown in in Table \ref{tab:Pass@k under different training strategies.}, the base model shown the \textit{pass@1} and \textit{pass@5} rates below 5\% and 16\%, respectively. In contrast, both the SFT model and the AssertSolver, which underwent pre-training and fine-tuning process with our augmented dataset, consistently achieved more than 80\% \textit{pass@1} and \textit{pass@5}. This presents a significant performance improvement, with over a 16-fold improvement in \textit{pass@1} and a 5-fold increase in \textit{pass@5}. 

In evaluating the performance between the SFT model and the AssertSolver, a clear differences is observed. The AssertSolver, which was further trained on errors from challenging cases, demonstrated an improvement in \textit{pass@1} performance, increasing from 84.66\% to 88.54\%. However, this was accompanied by a slight decline in \textit{pass@5} performance, from 91.64\% to 90\%, compared to the SFT model. As outlined in Section \ref{sec:Setup} evaluation metric, \textit{pass@1} and \textit{pass@5} reflect distinct characteristics of the underlying model. This observation suggests that although the model’s precision improves with additional training on errors from challenging cases, its ability to generate a diverse range of solutions decreases. 

Further analysis examined the performance of 915 test cases in the \textbf{SVA-Eval} dataset. In each cases, it generated \( n = 20 \) possible solutions. We evaluated these solutions, deeming \( c \) of them effective. These test cases were then categorised into 21 distinct outcomes ranging from `\( c = 0 \)’ (indicating no effective solutions) to `\( c = 20 \)’ (where all solutions were effective). Intermediate values suggested varying levels of success and correlated with increased uncertainty as shown in Fig.\ref{fig:Frequency distribution of correct answers across 20 responses.}. In the figure, the AssertSolver generally outperforms the SFT model in deterministic scenarios (i.e., `\( c = 0 \)’ and `\( c = 20 \)’) but fell short in non-deterministic ranges. This finding, as supported by Table \ref{tab:Pass@k under different training strategies.}, indicates that while adding challenging cases enhances the model’s precision, it may limit the diversity of the solutions.

\begin{table}[h]
\centering
\caption{Performance comparison between AssertSolver and other LLMs  (grey shading: the best performance across models).}
\label{tab:Performance comparison of models}
\fontsize{6.2}{11}\selectfont
\setlength{\fboxsep}{1.5pt}
\setlength{\tabcolsep}{1.3pt}
\renewcommand{\arraystretch}{1.0}
\resizebox{\columnwidth}{!}{
\begin{tabularx}{\columnwidth}{c c c c c c c c}
\toprule
\multirow{2}{*}{\textbf{Model}} 
& \multicolumn{2}{c}{\textbf{SVA-Eval-Machine}} 
& \multicolumn{2}{c}{\textbf{SVA-Eval-Human}} 
& \multicolumn{2}{c}{\textbf{SVA-Eval}} \\
\cmidrule(lr){2-3} \cmidrule(lr){4-5} \cmidrule(lr){6-7}
& \textit{pass@1 (\%)} & \textit{pass@5 (\%)} 
& \textit{pass@1 (\%)} & \textit{pass@5 (\%)} 
& \textit{pass@1 (\%)} & \textit{pass@5 (\%)} \\
\midrule

Claude-3.5
& 74.86 & 84.10 
& 66.58 & 77.48 
& 74.52 & 83.83 \\
GPT-4
& 58.04 & 78.45 
& 54.74 & 74.01 
& 57.90 & 78.27 \\
o1-preview 
& 76.96 & 87.73 
& 67.50 & \colorbox{gray!25}{87.94} 
& 76.57 & 87.74 \\
\midrule

Deepseek-coder-6.7b 
& 4.41 & 15.85 
& 2.89 & 10.27 
& 4.35 & 15.62 \\
CodeLlama-7b 
& 5.95 & 17.06 
& 4.47 & 12.85 
& 5.89 & 16.89 \\
Llama-3.1-8b
& 20.18 & 32.41
& 14.08 & 24.48
& 19.92 & 32.08 \\

\midrule

AssertSolver 
& \colorbox{gray!25}{89.04} 
& \colorbox{gray!25}{90.38} 
& \colorbox{gray!25}{76.97} 
& 81.35 
& \colorbox{gray!25}{88.54} 
& \colorbox{gray!25}{90.00} \\
\bottomrule
\end{tabularx}
}
\end{table}

\begin{figure}[t]
\vspace{-8pt}
\includegraphics[width=1\columnwidth]{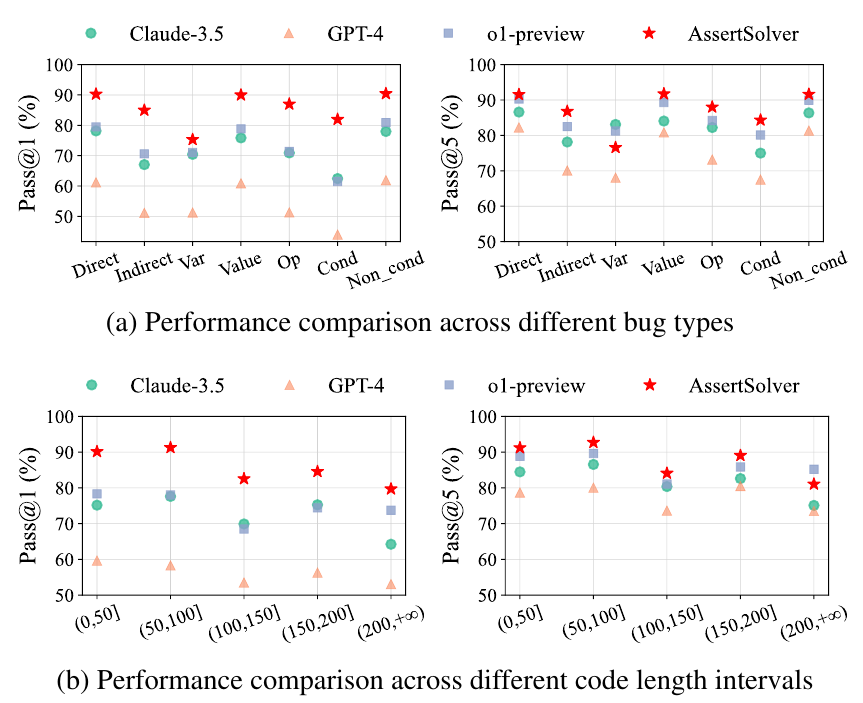}
\vspace{-12pt}
\caption{Comprehensive comparison with closed-source LLMs.}
\vspace{-15pt}
\label{fig:Comprehensive performance comparison}
\end{figure}

\noindent{\textbf{RQ2:}} For this research question, we compared the performance of AssertSolver with leading commercial and open-source LLMs, including the recently released o1-preview from OpenAI, focusing on \textit{pass@1} and \textit{pass@5} metrics within the \textbf{SVA-Eval} dataset. These comparisons are detailed in Table \ref{tab:Performance comparison of models}. AssertSolver outperforms all other models, achieving more than 80\% in the \textit{pass@1} metric (reaching 88.54\%) and 90\% in \textit{pass@5}, while the second-best model, the o1-preview, scored 76.57\% and 87.74\% respectively. Further analysis, divided by the methods used to generate bugs, indicated that AssertSolver consistently performs best in all categories, except for \textit{pass@5} in the \textbf{SVA-Eval-Human}. Considering the results from RQ1, where the AssertSolver demonstrated a preference for precision over diversity—thereby trading off performance in \textit{pass@1} for \textit{pass@5}—these outcomes align with expectations.

\noindent{\textbf{RQ3:}} As shown in Table \ref{tab:Performance comparison of models}, we observed that the performance for both \textit{pass@1} and \textit{pass@5} in \textbf{SVA-Eval-Human} consistently underperforms compared to \textbf{SVA-Eval-Machine}, with the exception of the \textit{pass@5} metric on the o1-preview model. Across all tested models, there was an average relative decline of approximately 19\% in \textit{pass@1} and 15\% in \textit{pass@5}, which were calculated by averaging the ratio of the \textit{pass@1} and \textit{pass@5} rates between the \textbf{SVA-Eval-Machine} and \textbf{SVA-Eval-Human} datasets. This observation suggests there may be a systemic difference between machine and human-generated bugs, but requiring further investigation to confirm.

\noindent{\textbf{RQ4:}} To answer this research question, we evaluated AssertSolver's performance across different bug types and code lengths against closed-source LLMs, as shown in Fig. \ref{fig:Comprehensive performance comparison}. AssertSolver consistently surpasses the performance of compared LLMs in \textit{pass@1} across all tested scenarios and outperforms in \textit{pass@5} in 10 out of 12 (83\%) scenarios. Despite slight underperformance in the `\textit{Var}' and in code length exceeding 200, AssertSolver demonstrates better results in all remaining cases. Particularly, for shorter code (under 100 lines) and the bugs classified as `\textit{Direct}', `\textit{Value}', and `\textit{Non\_cond}', both \textit{pass@1} and \textit{pass@5} reached over 90\%, showcasing AssertSolver's effectiveness across various design scenarios.

\begin{figure}[t]
    \centering
    \vspace{-2pt}
    \scalebox{.95}{\includegraphics[width=\columnwidth]{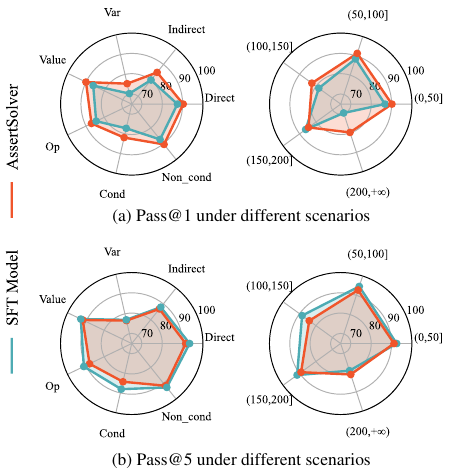}}
    \caption{The performance of STF Model and AssertSolver in various bugs types and code length intervals.}
    \label{fig:comparation of STF model and AssertSolver in various bugs types and code length intervals.}
    \vspace{-15pt}
\end{figure}

Further analysis, as shown in Fig. \ref{fig:comparation of STF model and AssertSolver in various bugs types and code length intervals.}, highlights how learning from error responses to challenging cases improves the \textit{pass@1} across nearly all scenarios, particularly for code lengths exceeding 200 lines, with the exception of code between 150-200 lines. While there is a slight decrease in \textit{pass@5} in these instances, this drop is not unexpected. AssertSolver prioritises precision over diversity, leading to performance that inherently favours \textit{pass@1} results.

\vspace{-2.5pt}
\section{Conclusion}
\label{sec:Conclusion}
\vspace{-2.5pt}

We presented AssertSolver, the first open-source LLM  designed to address assertion failures in RTL design. To overcome the challenge of data underrepresentation in training datasets, we implemented a data augmentation method that automatically enriches the training data with diverse assertion failure scenarios. Also, we introduced a novel training strategy that enables the model to learn not only from the augmented data but also from errors in challenging cases, thereby enhancing its capability to solve assertion failures effectively. 

Our experimental results show that AssertSolver achieves \textit{pass@1} and \textit{pass@5} rates of 88.54\% and 90.00\%, respectively, on a comprehensive test set of over 900 instances. This performance surpasses the recently released o1-preview by 11.97\% and 2.26\% in the \textit{pass@1} and \textit{pass@5} metrics, respectively. Furthermore, by learning errors from challenging cases, AssertSolver exhibits increased \textit{pass@1} rate. This significant performance underscores the model’s suitability for applications that require high reliability and consistency, such as solving assertion failures in hardware design. 

Moreover, we have made AssertSolver publicly available for early adoption and released an openly accessible evaluation benchmark for solving assertion failures, featuring various bug types generated by both machine and domain experts. Our work demonstrates that domain-specific fine-tuning of LLMs, combined with effective data augmentation and training strategies, can significantly advance the automation of solving assertion failures in hardware design.

\clearpage
\bibliographystyle{IEEEtran}
\bibliography{ref}

\end{document}